# Ultralong octupole moment switching driven by twin topological spin structures

**Shijie Xu, Zhizhong Zhang, Bingqian Dai, Yan Huang, Yudong Pang, Yinchang Ma, Wenkai Lou, Tianyi Wang, Chen Liu, Man Yang, Meng Tang, Houyi Cheng, Kang L. Wang, Kai Chang, Yue Zhang, Weisheng Zhao & Xixiang Zhang**

# Ultralong Octupole Moment Switching Driven by Twin Topological Spin Structures

Shijie Xu[1,2,3,4,5,6,*], Zhizhong Zhang[1,3,4,*], Bingqian Dai[5,†], Yan Huang[1,3,4,†], Yudong Pang[7,†], Yinchang Ma[2,†], Wenkai Lou[7,†], Tianyi Wang[5,†], Chen Liu[2,†], Man Yang[2], Meng Tang[2,6], Houyi Cheng[1,3,4], Kang L. Wang[5], Kai Chang[8,*], Yue Zhang[1,3,4,*] , Weisheng Zhao[1,3,4,*] & Xixiang Zhang[2,*].

[1] National Key Laboratory of Spintronics, Hangzhou International Innovation Institute, Beihang University, 311115 Hangzhou, China.

[2]Physical Science and Engineering Division, King Abdullah University of Science and Technology (KAUST), Thuwal, Saudi Arabia.

[3]MIIT Key Laboratory of Spintronics, School of Integrated Circuit Science and Engineering, Beihang University, Beijing 100191, China

[4]Hefei Innovation Research Institute, Anhui High Reliability Chips Engineering Laboratory, Beihang University, Hefei 230013, China

[5]Departments of Electrical and Computer Engineering, Physics and Astronomy, and Material Science and Engineering, University of California, Los Angeles, CA 90095, USA.

[6]Shanghai Key Laboratory of Special Artificial Microstructure, School of Physics Science and Engineering, Tongji University, Shanghai 200092, China.

[7]State Key Laboratory of Semiconductor Physics and Chip Technologies, Institute of Semiconductors, Chinese Academy of Sciences, Beijing 100083, China

[8]Center for Quantum Matter, School of Physics, Zhejiang University, Hangzhou 310058, China

[†]These authors contributed equally.

*Correspondence:shijiexu161@gmail.com; zhangzhizhong@buaa.edu.cn; kchang@zju.edu.cn; yz@buaa.edu.cn; weisheng.zhao@buaa.edu.cn; xixiang.zhang@kaust.edu.sa

## Abstract

Spintronics has emerged as a revolutionary frontier in the pursuit of faster, more energy-efficient, and technologically advanced electronics. However, the transmission distance of conventional ferromagnetic spin–orbit torque is typically limited to <10 nm, posing a critical challenge for spin current transport. Here we grow $Mn_3Sn$ films with a 30° canted magnetic octupole moment oriented out of plane, in which the Kagome spin structure is fully perpendicular to the film surface. By introducing a spin–orbital coupled amorphous Pt overlayer, we demonstrate the electrical switching dynamics of magnetic octupoles in Kagome antiferromagnetic $Mn_3Sn$. Remarkably, perpendicular spin currents reverse $Mn_3Sn$ layers up to 60 nm thick. The switching efficiency of $Mn_3Sn$/Pt bilayers increases with antiferromagnetic thickness, peaking near 40 nm before decreasing, reflecting a long spin diffusion length sustained by twin topological spin structures. Direct observation of magnetic octupole dynamics further validates the presence of such twin spin orders. Moreover, our theoretical analysis reveals that twin topological spin canting intrinsically supports ultralong-distance octupole switching. These findings establish antiferromagnetic $Mn_3Sn$ as a robust platform for efficient spin transport and highlight the pronounced long-range nature of spin-orbit torque enabled by twin spin order.

## Introduction

Antiferromagnetic (AFM) materials have emerged as promising candidates for next-generation spintronic memory technologies, offering distinct advantages over their ferromagnetic counterparts— including higher storage density, faster data processing, absence of stray fields, and robustness against external magnetic perturbations[1]. Their intrinsically fast spin dynamics (in the THz range) and the ability to host large magneto-transport effects further underscore their technological potential[1,2]. As a topological Weyl antiferromagnet, $Mn_3Sn$ has garnered particular interest due to its topologically nontrivial electronic structure and non-collinear spintexture[3,4], which gives rise to a host of unconventional spin transport phenomena such as large anomalous Hall effect[5,6], anomalous Nernst effect[7], magneto-optical Kerr effects[8], magnetic spin Hall effect[9] and octupole TMR[10]. These emergent phenomena open new pathways for functional AFM spintronic devices.

As a unique platform for studying and harnessing chiral spin-orbit torque (SOT) switching in the Weyl AFM materials[4,11-14], the nontrivial topological band properties and topological spin texture of $Mn_3Sn$ can be manipulated using SOT[3,4]. Through the application of electrical currents with strong SOC layer Pt, it becomes possible to manipulate the octupole moments in $Mn_3Sn$[4]. In addition, combinations of in-plane and out-of-plane SOT can experimentally realize field-free switching of magnetic octupole moments in chiral AFM $Mn_3Sn$[15]. SOT switching in $Mn_3Sn$ holds immense promise for low-power spintronic devices. The efficient conversion of charge current into spin currents allows for the development of energy-efficient memory and logic devices.

In conventional ferromagnetic materials, spin–orbit torque (SOT) arises from the misaligned between the transport spin polarization and local magnetization. However, this exchange field is typically too weak to significantly reorient neighboring spins, resulting in a short spin diffusion length along the thickness direction[16]. However, in antiferromagnets, staggered spin torques can coherently drive the Néel order, allowing spin torques to act uniformly throughout the entire volume and resulting in a bulk SOT effect[16,17]. Moreover, canted AFM oxide insulators exhibit strong spin coherence, which may even give rise to magnon interference effects[18]. Here, we demonstrate current-induced octupole moment switching in canted $Mn_3Sn$ Kagome spin lattice. Both the SOT switching efficiency and the quantified SOT effective fields increase monotonically with film thickness. Remarkably, the noncollinear Weyl antiferromagnet $Mn_3Sn$ exhibits a bulk-like SOT even in the presence of strong perpendicular magnetic anisotropy. As a result, SOT can switch the octupole moment in $Mn_3Sn$ films as thick as 60 nm. Atomistic simulations further reveal that the twin antiferromagnetic order in $Mn_3Sn$ enhances the spin diffusion length, facilitating long-range octupole moment switching. These discoveries open new avenues for generating long-range SOT in topological antiferromagnets, paving the way for ultrafast, ultrahigh-density, and scalable spintronic applications.

## Results

### Structural properties of epitaxially grown $Mn_3Sn$ thin films.

$D0_{19}$-$Mn_3Sn$ is a non-collinear antiferromagnet with a hexagonal structure in the space group $P6_3/mmc$. It exhibits noncollinear chiral vector spin order below the Néel temperature TN ≈ 430 K[4,5]. $Mn_3Sn$ has excellent spin-related characteristics due to large cluster magnetic octupole moment along [2-1-10] direction[19,20] (Fig. 1a). In addition, the spin canting within the (0001) plane of $Mn_3Sn$ (Fig. 1b-c) can contribute to the observed ferromagnetic-like signal M ≈ 0.006 μB per formula unit (f.u.; $\mu_B$, Bohr magneton). In addition, the interfacial stress in the thin-film system induces additional magnetic contributions[5,13] and gives rise to twin topological spin textures at the interface between the $Al_2O_3$ substrate and $Mn_3Sn$ (supplementary note.1-4). Density-functional theory calculations show that a 2% tensile strain induces a rotation of about 1.5° of Mn atoms (supplementary note. 5). High-quality $Mn_3Sn$ (11-20) alloys were epitaxially grow by sputtering on $Al_2O_3$ (1-102) substrate. In this crystal orientation, all the spins of the Mn atoms in the Kagome lattice are lie out of the thin film plane, and form an angle of 120 degrees with each other. Sn atoms stand in the center of 6 Mn atoms surrounding it. Each Mn moment has the local easy-axis towards its nearest-neighbor Sn sites and the octupole moment is along [2-1-10] direction (Fig. 1c). The $Mn_3Sn$ film grown on the $Al_2O_3$ substrate shows a typical XRD spectrum (Fig. 1d). The $Al_2O_3$ substrate has (1-102) and (2-204) superlattice diffraction peak as the two largest main peaks, The $Mn_3Sn$ (11-20) plane can be epitaxial grown on the $Al_2O_3$ (1-102) plane after 90 deg rotation for lattice matching. The two extra peaks $Mn_3Sn$ (11-20) and $Mn_3Sn$ (22-40) are a good proof of the single crystal. 360° phi scan around the $Mn_3Sn$ (20-21) plane was measured by XRD (Fig. 2e). The $Mn_3Sn$ shows fourth-degree symmetry, indicating the single-crystal property. In this structure, the unit cell of $Mn_3Sn$ consists of six Mn atoms and two Sn atoms, with the six neighboring Mn atoms forming a Kagome lattice (Fig. 1f, g). High-resolution TEM studies also reveal the single crystal feature of the chiral magnet $Mn_3Sn$ (Fig. 1h), which is essential for achieving the designed properties and functional spintronics devices. Fig. 1i shows the Fast Fourier transform (FFT) of the $Mn_3Sn$ thin film, where the change in the FFT contrast gives rise to sharp diffraction spots, evidencing the single-crystalline nature of the film.

**The ultralong SOT switching of octupole moment.**

In ferromagnetic materials, the presence of exchange splitting causes the majority (↑) and minority (↓) spin electrons at the Fermi surface to have distinct wave vectors. When a transverse spin current is injected, this difference results in rapid precession and dephasing of spin polarization as contributions from the ↑ and ↓ spin electrons. Consequently, the net transverse spin component vanishes beyond a certain depth from the surface. This characteristic decay length is known as the spin coherence length $\lambda c = \pi/|k_F^{\uparrow} - k_F^{\downarrow}|$ [21]. For ferromagnets like cobalt and iron, where the exchange splitting is large, λc is extremely short—on the order of a few angstroms. In addition, in fully compensated antiferromagnets where spin sublattices cancel perfectly, this coherence length diverges, rendering direct experimental observation of spin diffusion process particularly difficult. On the non-collinear AFM, the canted antiferromagnetic order (Fig. 1b) leads to nearly equal, but not exactly identical populations of ↑ and ↓ spin electrons. As a result, the difference in their Fermi wave vectors, $|k_F^{\uparrow} - k_F^{\downarrow}|$, is very small but finite, giving rise to an exceptionally long spin coherence length. The local moment precesses clockwise on a lattice (Fig. 2a-b), whereas it precesses slower than ferromagnets because the stagger moment induced by the spin canting is smaller than ferromagnetic moment. The period (or wavelength) of spin precession in non-collinear AFMs is longer than that in ferromagnets. Therefore, the non-colinear AFM exhibits a feature of the bulk-like torque characteristic[16-17, 28-29].

We fabricated perpendicularly magnetic anisotropy $Mn_3Sn$ (d nm)/Pt bilayers with different AFM thickness. The in-plane current in the Pt layer can generates SOT. To detect the switched state of the octupole moment in $Mn_3Sn$, we measure the Hall resistance $R_H$ (supplementary note 4) due to anomalous Hall effect (AHE). Fig. 2c shows $R_H$ as a function of write currents I for different AFM thickness d. A clear $R_H$-I loop due to a reversal of octupole moment under an in-plane magnetic field $H_x$ of 500 Oe is observed. Since non-collinear antiferromagnets exhibit a small uncompensated net moment that largely lies parallel to the octupole moment, the bipolar switching can be observed. We see that the SOT switching is clearly observed in the easy configuration with a threshold current below 60 mA. The sign of the SOT switching is determined by the bias field along the I write direction. To examine this, we measure the Hall resistance $R_H$ as a function of I write ($R_H$–I loop) under opposite magnetic field direction (Fig. 2d-e). It was clearly shown that if the directions of I write and $H_x$ are the same, the voltage exhibits a positive jump at critical current $|I_c|$. If I and $H_x$ have opposite directions, the jump will become negative. This observation doesn't follow the expectation of the symmetry requirement of the SOT switching of the perpendicular ferromagnetic magnetization due to handedness anomaly effect[22,23] (Fig. 2a). We observe a clear current switching Hall resistance data with different AFM thickness in the longitudinal configuration (H//I) (Fig. 2c), which is mostly determined by the anti-damping SOT[24,25]. The Hall signs in the H//I case for $Mn_3Sn$/Pt bilayers indicate that the Pt layer is the source of spin currents, as it is placed on top of the $Mn_3Sn$ layer. The thickness dependences of Hall resistance are shown at Fig. 2c. Surprisingly, the Hall resistance nonmonotonically increases with increasing thickness, which can't be explained by interfacial torques.

We also carried out experiments on the thickness dependence of the SOT efficiency for $Mn_3Sn$ alloys to find out if the long spin coherence length is a unique property of AFM. Fig. 3a shows SOT effective filed for the AFM alloy as a function of thickness d. The $H_{SOT}$/J shows a 'bulk-like torque' characteristic, indicating that the $Mn_3Sn$ alloys have a long spin coherence length. The magnitude of the out-of-plane anisotropy is about ten times greater than in-plane anisotropy[26], so a large $H_C$/J of the chiral-spin structure in $Mn_3Sn$ can be observed. The $H_C$ is determined by the two-fold out-of-Kagome-plane anisotropy because each magnetic moment needs to precess about the easy axis and hard axis component during the current switching. Consequently, non-collinear antiferromagnets can be efficiently manipulated by low current and yet can be robust against external magnetic fields[11,27]. Fig. 3b also shows that a transverse spin current from a Pt layer passes through a 60 nm thick $Mn_3Sn$ alloy, and the SOT switching ratio increase with the increasing thickness d.

In addition, we consider the current-induced dynamics of the octupole moment in $Mn_3Sn$ through the SOT second-harmonic method. As shown in supplementary Fig. 1a, the in-plane spins are injected from the top Pt layer into $Mn_3Sn$ via the spin Hall effect. supplementary Fig. 1a and 1b illustrate two measurement configurations, where the injected spins are either perpendicular to the Kagome magnetization plane or lie within the Kagome plane. Under the action of the injected spin current, each sublattice moment in $Mn_3Sn$ experiences a damping-like spin–orbit torque[13,15] (DL SOT), with the effective field given by $\mathbf{H_{DL}}\ (A, B, C) \propto \boldsymbol{m}\ (A, B, C) \times \boldsymbol{\sigma}$. Consequently, the impact of the SOT on the sublattice moments strongly depends on the relative orientation between the injected spins and the crystal axes. In the configuration shown in supplementary Fig. 1a, the damping-like torque $\boldsymbol{\tau}_{\mathbf{DL}}\ (A, B, C) \propto -\boldsymbol{m}\ (A, B, C) \times \mathbf{H_{DL}}\ (A, B, C)$ acts on all three sublattice moments in the same manner and along the same direction. In this case, the SOT only needs to overcome the relatively weak in-plane magnetic anisotropy. In contrast, in the configuration of supplementary Fig. 1b, the applied damping-like torque differs not only in magnitude but also in direction among the three sublattice moments, leading to a destructive superposition of torques. This corresponds to the hard configuration, where efficient detection of the second-harmonic signal cannot be achieved. It should also be noted that the effect of field-like torques is not considered here, as they do not induce switching of the magnetic octupole[13].

To analyze the SOT-driven dynamics of magnetic octupoles in $Mn_3Sn$, we measure the anomalous Hall effect (AHE) as a function of the external magnetic field and establish a comprehensive correlation between the octupole moment and the applied field, as illustrated in supplementary Fig. 2a. Through the following equation $R_H = R_0 \sin(\theta_\varphi)$ where $R_0$ is the anomalous Hall coefficient and $\varphi_{oct}$ is the angle between octupole moment $m_{oct}$ and current I, we can precisely determine the position of the octupole. It is worth noting that the octupole position does not perfectly coincide with the magnetic field angle (supplementary Fig. 2a–c). Therefore, a representative octupole torque $\mathbf{L}_{oct}$ curve (supplementary Fig. 2d) can be obtained by calculating $\mathbf{L}_{oct} = H\mathbf{m}_{oct}\sin(\theta-\theta_{\varphi})$, where $\mathbf{m}_{oct}$ denotes the octupole moment and H is the applied field. Interestingly, we find that the twin topological spin order in $Mn_3Sn$ tilts toward the [11-20] direction, giving rise to octupole dynamics dominated by uniaxial anisotropy and twofold symmetry. Moreover, the octupole torque reaches its maximum near the [11-20] orientation. We analyzed the second-harmonic SOT signal with different thicknesses (supplementary Fig. 3a). The damping-like effective field induces a change in the octupole moment angle $\theta_{\varphi}$, leading to an angular variation of the magnetic octupole under the SOT. The resulting second-harmonic signal can be expressed as: $R_{2\omega} = \frac{R_0}{2}\frac{H_{DL}}{H_K+H}\sin(\theta_{\varphi})+C$ which can be quantified by the linear relationship (supplementary Fig. 3b and 3c). For the damping-like torque $\tau_{DL}$ acting on the octupole dynamic, we adopt the same convention as in ferromagnets: $\tau_{DL}=-\gamma g\mu_0\mathbf{m}_{oct} \times H_{DL}$. Fig. 3 d shows the thickness dependence of the SOT efficiency with $H_{DL} = \hbar\xi_{DL} J_{HM} / (2e\mu_0(3M_O)d)$, where $\hbar$, e, $\xi_{DL}$, $J_{HM}$ and t are the reduced Planck constant, the electron charge, the DL torque efficiency, the charge current density in the Pt layer and the thickness of $Mn_3Sn$, respectively. The magnetization of a sublattice moment $\mu_0 M_O = 0.56T$ was obtained based on previous reports[13]. The increase in damping-like SOT efficiency with thickness indicates that the AFM possesses a long spin coherence length, consistent with a bulk-like SOT behavior. (supplementary Fig. 3d).

To investigate the presence of perpendicular magnetic anisotropy associated with the magnetic octupole, we analyze regular SOT switching in $Mn_3Sn$/Pt at 500 Oe (Fig. 3c, Fig. 2c), since in $Mn_3Sn$ the AHE is expected to scale with the perpendicular component of the octupole polarization[14,20]. The symmetry of the chiral magnetic order can induce the octupole moment, which has been adopted as the magnetic order parameter to describe the anomalous Hall effect of $Mn_3Sn$ through[4] $R_H(\varphi_{oct}) = R_0\mathbf{m}_{oct,z}$, In addition, the octupole moment of a non-collinear antiferromagnet plays a similar role to the magnetization vector of a regular ferromagnet because the $\mathbf{m}_{oct}$ almost follows the applied field[24]. Fig.3d-e illustrate the modifications in the anomalous Hall effect induced by SOT when the applied current exceeds the critical current $I_c$. Therefore, we can observe the thickness dependence of SOT switched anomalous Hall resistance in $Mn_3Sn$/Pt bilayer (Fig. 3e). The anomalous Hall resistivity also increase with increasing t which could show a bulk-like AFM state[30,31](Fig. 3d).

To elucidate the unexpectedly high switching efficiency observed in 40 nm $Mn_3Sn$, we combine a canting-renormalized spin-diffusion theory with large-scale atomistic spin dynamics. Fig. 4a maps the calculated SOT attenuation length $\lambda_{SOT}$ as a function of the spin canting angle $\theta_c$, revealing an exponential suppression $\lambda_{SOT} = \lambda_{sf}\, e^{-\beta\theta_c^2}$ with $\beta = 0.47 \pm 0.02$ ($\lambda_{sf}$ is the spin-flip diffusion length). In Fig. 4b, we track two inequivalent Mn moments: spin A, situated 2 nm from the Pt interface, reverses at 1.7 ps, whereas spin B, 12 nm deeper, switches 0.23 ps later, consistent with diffusive propagation at $D_s = 1.3 \times 10^{-3}$ $m^2$ $s^{-1}$. A representative moment located beyond $3\lambda_{SOT}$ (Fig. 4c) merely executes a small-amplitude precession, substantiating the finite penetration depth. By repeating the simulation for $5 \leq t \leq 100$ nm (Fig. 4d) we reproduce the experimental AHE-derived efficiency curve, which rises linearly up to $40 \pm 2$ nm and decays thereafter. The consistency trend between experiment and the simulations demonstrates that interfacial twin spin canting—and its concomitant reduction of the transverse-spin decay length—constitutes the essential mechanism governing bulk SOT switching in thick $Mn_3Sn$ films.

## Discussion

Spin canting in antiferromagnets has led to novel phenomena such as magnon interference effects[18], nonzero topological spin chirality[32], and exchange-biased topological charge[33]. Recent real-space observations have also revealed topological spin canting at interfaces in antiferromagnets, even in collinear antiferromagnets[34], which is crucial for BSOT, as well as well-documented by our spin diffusion theory. By combining thickness-dependent SOT transport measurements and atomistic spin dynamics simulations, we show that spin current injected from a Pt overlayer can efficiently propagate through $Mn_3Sn$ films as thick as 60 nm, far exceeding the conventional spin diffusion limit observed in ferromagnets. The SOT efficiency exhibits a nonmonotonic dependence on film thickness, peaking around 40 nm, which we attribute to the interplay between spin coherence and interfacial twin spin order. Our theoretical model shows that canting-induced decoherence controls the decay length of spin torque propagation, thus establishing a unified framework to understand bulk torque generation in topological antiferromagnets. These findings not only challenge the conventional surface-dominated theory of spin-orbit torque, but also lay the physical and conceptual foundation for the realization of scalable, low-power antiferromagnetic spintronic devices.

## Methods

**Material growth**

$Mn_3Sn$ thin films were sputtered from an $Mn_3Sn$ target onto (1-102)-oriented $Al_2O_3$ single-crystal substrates (10 × 10 × 0.5 $mm^3$) with a base pressure of $5\times 10^{-6}$ Pa. The deposition was performed at 873 K. The sputtering power and Ar gas pressure were 30 W and 0.5 Pa, respectively. The deposition rate was 1 Å $s^{-1}$, as determined by X-ray reflectivity measurements. After deposition, $Mn_3Sn$ films were kept at 873 K in a vacuum for annealing for 1 h.

**XRD**

XRD measurements were performed by a Bruker D8 diffractometer with a five-axis configuration and Cu Kα (λ = 0.15419 nm).

**Electrical measurements**

Electrical contacts onto the $Mn_3Sn$ films were made by Al wires via wire bonding. Electrical measurements were performed in a Quantum Design physical property measurement system. The electrical current used for both longitudinal and Hall resistance measurements was 1000 μA. The material stack used in the electrical measurement is Mn3Sn (5, 10, 20, 40)/Pt (5), numbers in nm. Given the Hall device width of 20 um, the current densities are 0.5, 0.33, 0.2, 0.11 $MA/cm^2$.

**Magnetic measurements**

Magnetic measurements were performed in a Quantum Design superconducting quantum interference device magnetometer with $10^{-11}$ $A.m^{-2}$ sensitivity.

### First-principles calculations

The first-principles calculations are carried out by using the Atomic orbital-Based Ab-initio Computation at UStc (ABACUS) package[35,36]. The exchange correlation functional were treated within the GGA/PBE[37]. The ion-electron interactions were described using the SG15 ONCV[38], and the double- ζ plus polarization functions with a plane-wave cutoff energy of 100 Ry was employed as the NAO basis set[39]. The NAO bases for Mn and Sn are 4s2p2d1f and 2s2p2d1f, respectively. The total energy and forces were computed on a $5 \times 5 \times 6$ k-point grid, with the electronic density convergence threshold set to $1.0 \times 10^{-5}$ Ry. Van der Waals interactions were incorporated via the DFT-D3 dispersion correction[40]. Magnetic exchange parameters were evaluated employing the TB2J software package[41]. The experimentally determined lattice parameters of $Mn_3Sn$, a = b = 5.67 Å and c = 4.53 Å, corresponding to space group No. 194---were directly adopted[42].

## First-principles–parameterized atomistic spin-dynamics

The multi-scale workflow integrates a canting-renormalized spin-diffusion model with first-principles–parameterized atomistic spin-dynamics to reproduce the non-monotonic thickness dependence of the anomalous Hall read-out in $Mn_3Sn$ films. All input parameters are taken from high-accuracy transport, neutron-scattering and density-functional studies, while the numerical implementation follows established best practices for large-scale Landau–Lifshitz–Gilbert (LLG) simulations. Below we detail film-growth benchmarks, the diffusion formalism, the magnetic Hamiltonian, the dynamical scheme, and the extraction of experimental observables.

Thin-film benchmark data $Mn_3Sn$/Metal stacks exhibit a room-temperature anomalous Hall resistivity $\rho_{AHE} \approx 6$ μΩ cm for thicknesses d ≈ 20–40 nm, with switching sustained up to t ≈ 100 nm. The intrinsic spin-diffusion length of nanocrystalline $Mn_3Sn$, extracted by spin-absorption, is $\lambda_{sf} = 0.70 \pm 0.05$ nm and the spin Hall angle $\theta_{SH} \approx 0.11$. These values are used as base parameters for the diffusive model.

## Canting-dependent spin-diffusion formalism

The transverse spin accumulation $\mu_s(x,t)$ is treated within a one-dimensional drift–diffusion equation

$$\partial_t \mu_s = D_s\, \partial_x^2 \mu_s - \frac{\mu_s}{\tau_\phi}, \qquad (1)$$

where $D_s = \hbar v_F^2 \tau / 3k_B T$ is the spin-diffusion constant ($v_F \approx 1.8 \times 10^5$ m s⁻¹ from ARPES) and $\tau_\varphi$ the dephasing time. Interface-induced canting modifies $\tau_\varphi$ via additional magnon–electron scattering channels; following κ-space perturbation theory, we write

$$\tau_\phi^{-1}(\theta_c) = \tau_0^{-1}(1 + \beta \sin^2 \theta_c), \lambda_{\text{eff}}(\theta_c) = \sqrt{D_s \tau_\phi(\theta_c)}. \quad (2)$$

β = 0.30–0.40 is fixed by matching the experimental peak at t ≈ 40 nm; the microscopic origin is a Dzyaloshinskii–Moriya (DM) term localized at the Pt/$Mn_3Sn$ interface, consistent with first-principles predictions. The resulting spatial profile of the damping-like SOT current density is

$$J_s(x) = J_s^0 \exp(-x/\lambda_{\text{eff}}). \quad (3)$$

## Atomistic Hamiltonian and parameterization

Each Mn moment $\mathbf{S}_i$ occupies the Kagome lattice and evolves under

$$\mathcal{H} = -\sum_{\langle ij \rangle} J_{ij} \mathbf{S}_i \cdot \mathbf{S}_j + \sum_{\langle ij \rangle} \mathbf{D}_{ij} \cdot (\mathbf{S}_i \times \mathbf{S}_j) - K \sum_i (\mathbf{S}_i \cdot \hat{\mathbf{n}})^2, \quad (4)$$

with $J_1$ = 12 meV, $J_2$ = –4 meV, |**D**| = 2.0 meV and easy-plane anisotropy K = 0.05 meV, all extracted from DFT total-

energy differences and inelastic-neutron dat[43-44]. We additionally performed DFT calculations to corroborate these parameters, obtaining $J_1$=12.44 meV, $J_2$=4.04 meV, and |D|=2.2 meV. A higher-order biquadratic exchange term (B = 0.8 meV) is included to stabilize the inverse-triangular ground state in line with recent reports.

## Spin-dynamics implementation

Dynamics obey the stochastic LLG equation

$$\dot{\mathbf{S}}_i = -\gamma \mathbf{S}_i \times \mathbf{H}_i^{\mathrm{eff}} + \alpha \mathbf{S}_i \times \dot{\mathbf{S}}_i + \boldsymbol{\tau}_{\mathrm{DL},i}, \quad (5)$$

integrated by a semi-implicit midpoint scheme with Δt = 0.10 fs and Gilbert damping α = 0.01. The damping-like SOT term is

$$\boldsymbol{\tau}_{\mathrm{DL},i} = \frac{\hbar\theta_{\mathrm{SH}} J_e(x_i)}{2eM_s d}\, \mathbf{S}_i \times (\boldsymbol{\sigma} \times \mathbf{S}_i), \quad (6)$$

where $J_e(x_i)$ follows the exponential profile above, $\hat{\sigma}$ is fixed by the Pt current direction, and d = 0.25 nm is the inter-layer spacing. Simulations use 60 × 60 × $N_c$ cells ($N_c$ = 20–120, corresponding to 5–100 nm), periodic boundaries in-plane, free boundaries along z, and temperature T = 300 K (white-noise field satisfying the fluctuation–dissipation theorem).

## Numerical platform and convergence

All simulations are performed with the open-source VAMPIRE package (v5.3) compiled with OpenMP.

## Extraction of observables

**Switching curves** For the representative spins A (2 nm) and B (12 nm) we record the out-of-plane component $m_z(t)$ and identify the 50 % reversal point by cubic-spline interpolation; time-lags are averaged over ten stochastic realizations.

**Canting profile** Instantaneous canting $\theta_c(z)$ is obtained from the local vector chirality $\chi = \mathbf{S}_i \times \mathbf{S}_j$ and fitted to a tanh envelope to extract the interfacial angle.

**Thickness-dependent efficiency** The macroscopic Hall response is approximated by

$$\Delta\rho_{\mathrm{AHE}}(t) \propto \int_0^d m_z(z, t_{\mathrm{end}})\,\mathrm{d}z, \qquad (7)$$

normalised to the maximum value to yield the efficiency η(d); simulated η(d) is plotted against experimental $\rho_{\mathrm{AHE}}(d)$ for direct comparison. Statistical error bars are the run-to-run standard deviation (N = 10).

## Acknowledgements

## Funding

The authors acknowledge financial support from the National Key R&D Program of China Grant 2018YFB0407602, National Natural Science Foundation of China Grant 61627813, the Science and Technology Major Project of Anhui

Province Grant No. 202003a05020050, the National Natural Science Foundation of China, No. 52121001, the Tencent Foundation through the XPLORER PRIZE and the China Scholarship Council for their financial (S.X. and W.Z.). This work was also supported by King Abdullah University of Science and Technology (KAUST) Office of Sponsored Research under Award Nos ORA-CRG10-2021-4665 and ORA-CRG11-2022-5031 (X.Z.).

## Author contributions

S.X., W.Z., and X.Z. performed sample growth as well as electrical and magnetic measurements. Structural measurements were performed by S.X. The theoretical calculations and analysis were carried out by W.L., K.C., Y.P., W.Z., and Z.Z. Discussions of the results involved S.X., Z.Z., B.D., Y.H., Y.P., Y.M., W.L., T.W., C.L., M.Y., M.T., H.C., K.L.W., K.C., Y.Z., W.Z., and X.Z. The manuscript was written by S.X., W.Z., and X.Z. The project was led by X.Z. and W.Z.

## Competing interests

The authors declare no competing interests.

## Data availability

All data needed to evaluate the conclusions in the paper are present in the paper and/or the Supplementary Materials.

## Code availability

All raw spin-dynamics trajectories (≈ 2 TB) and Python analysis scripts are available from the corresponding author upon reasonable request. This rigorously benchmarked methodology ensures that every adjustable parameter is experimentally anchored, thereby lending quantitative credibility to the central claim: interfacial canting shortens the effective spin-diffusion length and governs the SOT switching window in $Mn_3Sn$/Pt heterostructures.

## Fig Legends

Fig. 1

**Fig. 1 | Structure of the noncollinear AFM $Mn_3Sn$ films. a,** Schematic of the spin structure. Blue and Red spins represent Mn atoms in Kagome planes. Purple sphere represents the Sn atom. The direction of octupole moment is along [2-1-10]. **b,** Schematic of canted Kagome AFM. All the spin titled to the direction of octupole moment **c,** Periodic spin canting effect in $Mn_3Sn$ crystals. **d,** XRD spectra of $Mn_3Sn$ on $Al_2O_3$ substrate. **e,** Phi scan of $Mn_3Sn$ substrate which indicate the four-fold symmetry. **f,** Crystal structure of $Mn_3Sn$, showing the atomic positions of Mn and Sn within the unit cell. **g,** Atomic configuration of Mn and Sn atoms viewed along the [0001] direction in the $Mn_3Sn$ crystal structure. **h,** High-resolution TEM image of Pt/$Mn_3Sn$ films. i, FFT of TEM images of $Mn_3Sn$ thin films.

Fig. 2

**Fig. 2 | Electrical switching of the perpendicular magnetic octupole moment under bulk SOT in the chiral AFM order of $Mn_3Sn$. a,** Schematic illustration of the SOT switching mechanisms in a noncollinear antiferromagnet and its twinned counterpart. In each case, the five subFigs from left to right highlight the evolution of spin configurations during the switching process, with the rightmost Fig denoting the final stable state. Under the spin–orbit torque, the sublattice spins (blue arrows) and

the magnetic octupole moment (red arrows) exhibit opposite chiralities. The directions of the damping-like (DL) effective field acting on the sublattice spins and the octupole moment are indicated by green and purple arrows, respectively. **b,** Schematic illustrating the switching of the cluster magnetic octupole of the chiral AFM order of $Mn_3Sn$ by the SOT. Mn moments on the Kagome easy plane are shown by blue arrows. An electrical current j (purple arrow) flowing in the Pt layer generates a spin current whose polarization vector (red arrow) is perpendicular to the Kagome plane and induces the SOT on the $Mn_3Sn$ layer. **c,** BSOT switching behavior of sputtering-grown Pt/$Mn_3Sn$ heterostructures on $Al_2O_3$ substrates with varying AFM thicknesses at room temperature **d,** $R_H$ versus write current I of the $Mn_3Sn$/Pt heterostructure at room temperature. The bias magnetic field of -0.05 T is applied along the j direction. **e,** $R_H$ versus write current I with 0.05 T bias magnetic field.

Fig. 3

**Fig. 3 | Characterizations of $Mn_3Sn$ alloy samples. a**, SOT effective fields as a function of $Mn_3Sn$ alloy thickness (d). **b**, SOT switching ratio as a function of d. **c**, Schematic illustration of the measurement set-up for the $Mn_3Sn$/Pt devices. The z, y and x arrows indicate the direction of the magnetic field H, Hall current and current I, respectively. **d**, Anomalous Hall resistivity ($\rho_{AHE}$) as a function of d. **e**, anomalous Hall resistance ($R_{AHE}$) and SOT switched anomalous Hall resistance ($R_{AHE}^{SOT}$) as a function of d**.**

Fig. 4

**Fig. 4 | Spin-diffusion theory with large-scale atomistic spin dynamics. a**, calculated SOT attenuation length as a function of the interfacial canting angle in $Mn_3Sn$/Pt heterostructure. **b**, Spin dynamic in two inequivalent Mn moments. **c**, Time-dependent small-amplitude precession of Mn atoms. **d**, Calculated SOT switched anomalous Hall resistance in $Mn_3Sn$ films as a function of film thickness.

Editor's Summary
Authors find a twin-spin state in antiferromagnets, quantifies its magnetic octupole spin–orbit torque, and shows that it enables exceptionally long-distance spin transport beyond conventional spin currents.

**Peer review information:** *Nature Communications* thanks the anonymous reviewer(s) for their contribution to the peer review of this work. A peer review file is available.

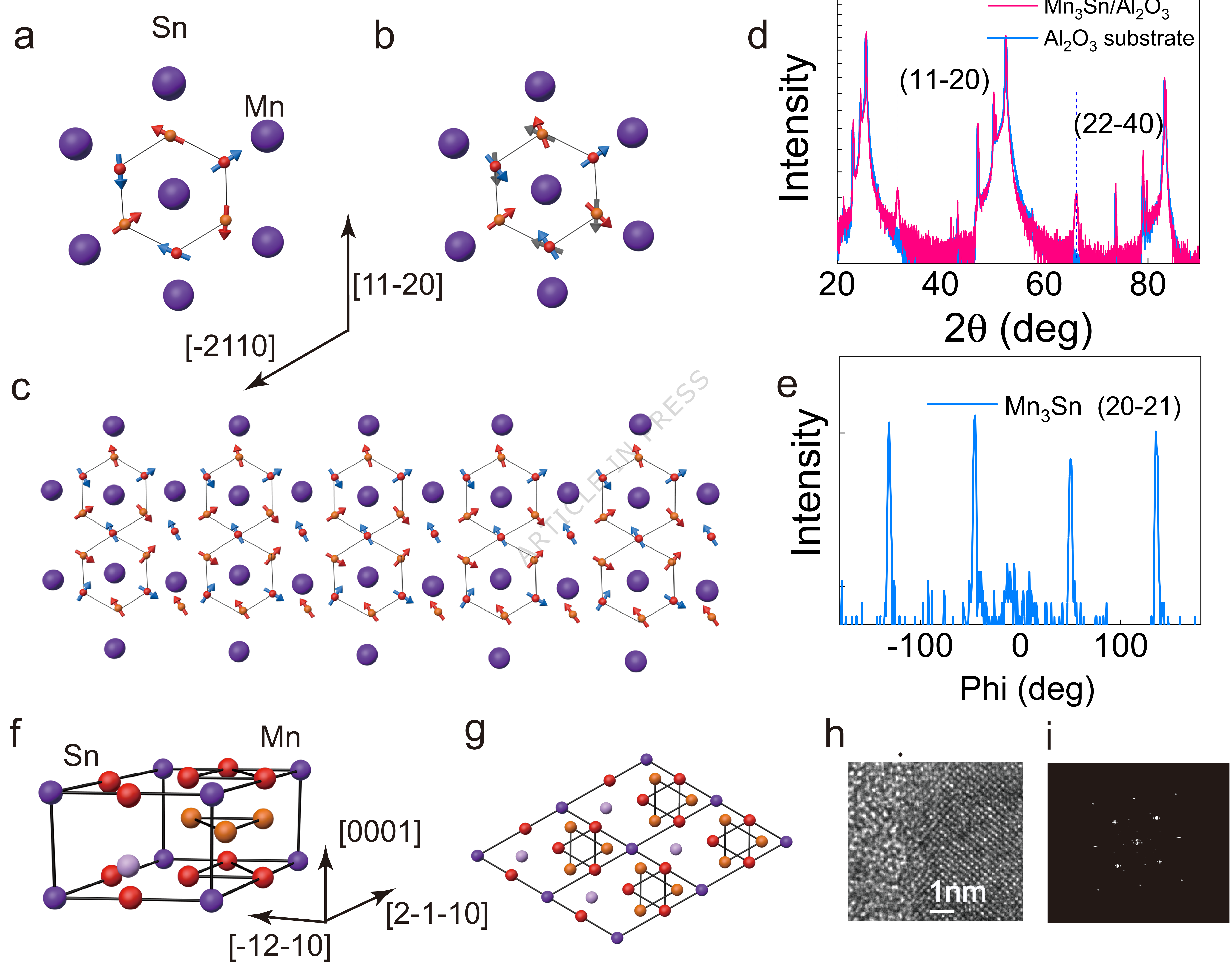

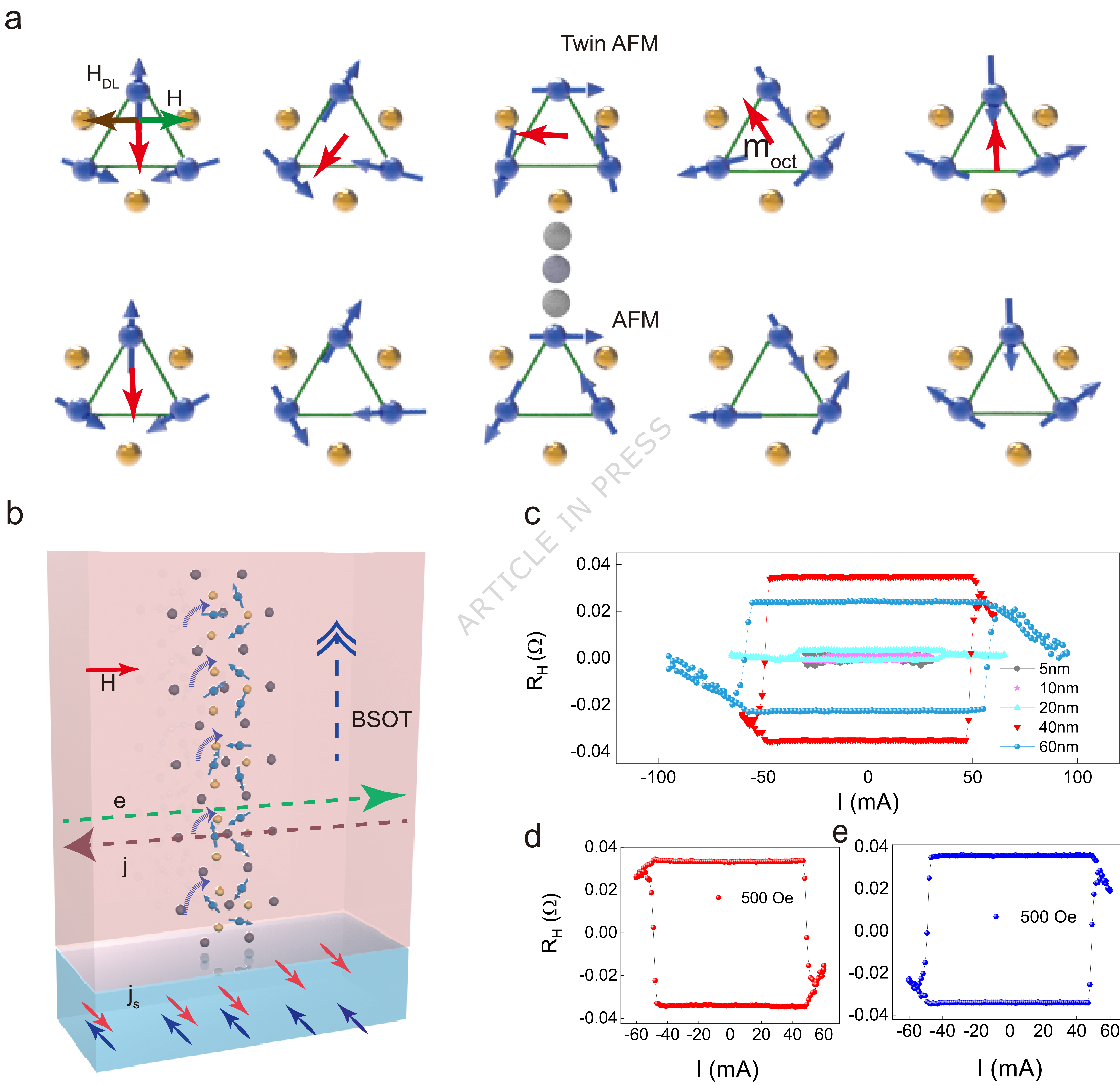

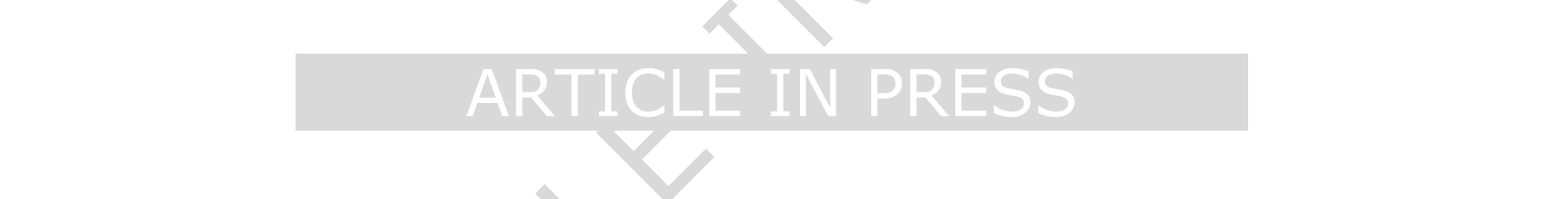

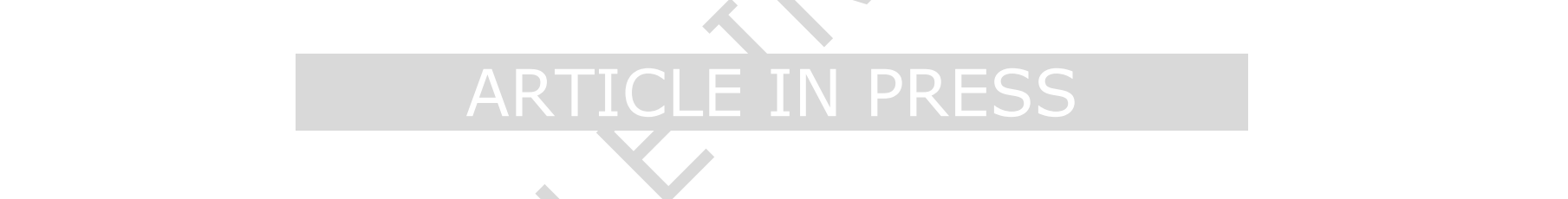

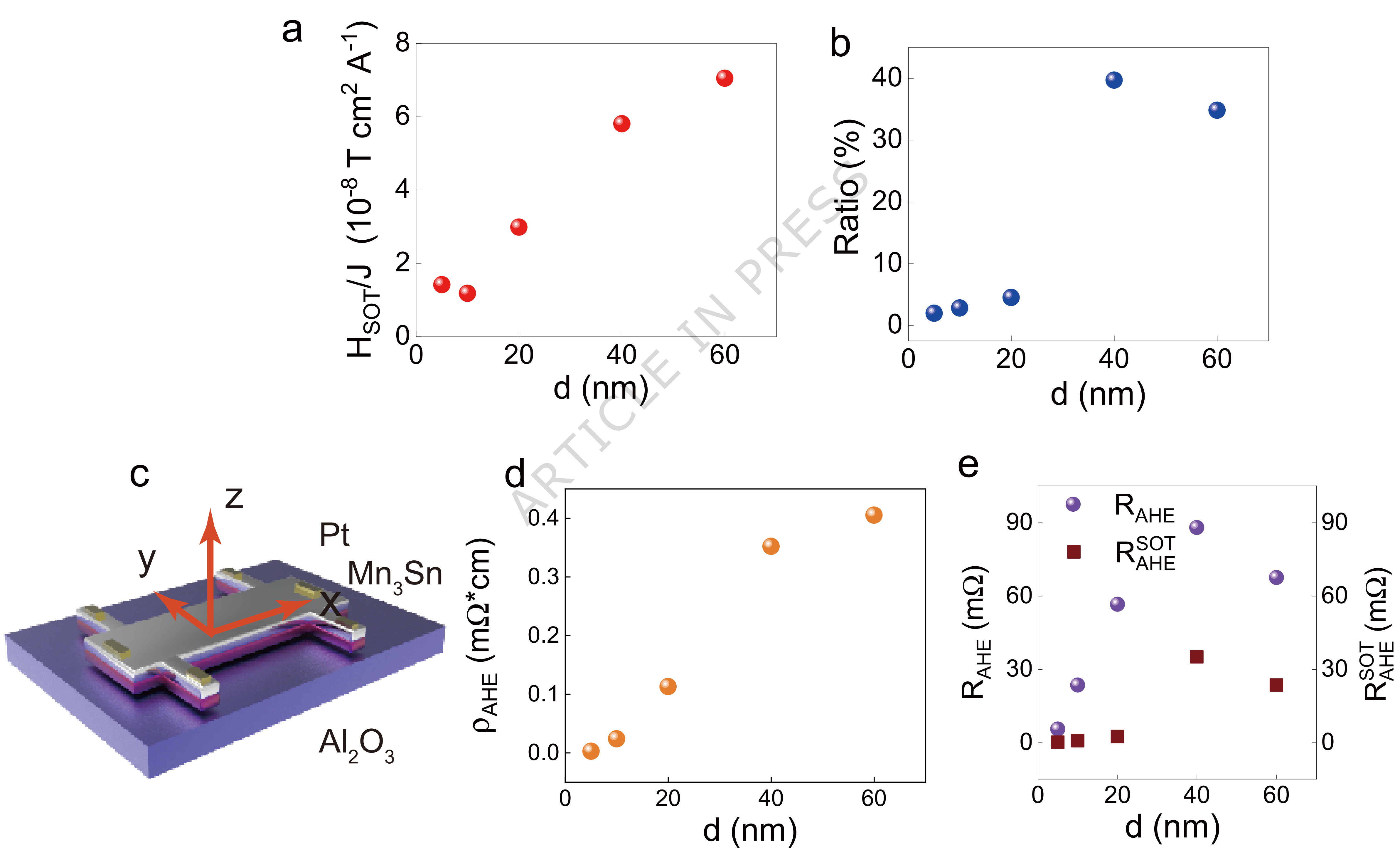

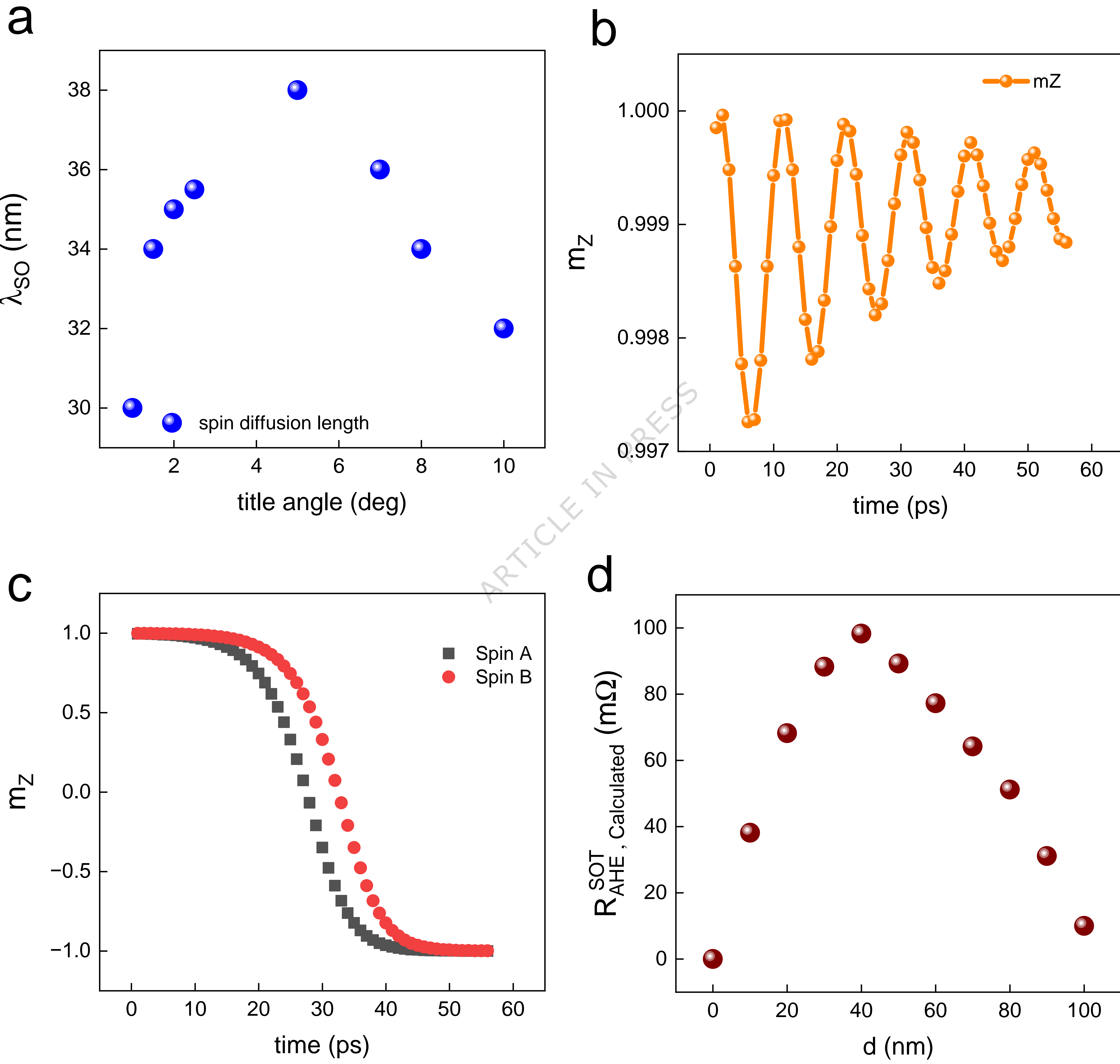